\documentclass[twocolumn,preprintnumbers,amsmath,amssymb]{revtex4}

\usepackage{dcolumn}
\usepackage{amstext}
\usepackage{amsmath}
\usepackage{amssymb}
\usepackage[pdftex]{graphicx}
\usepackage{subfigure}
\usepackage[T1,OT1]{fontenc} 

\begin{document}
\title{Soft elastic surfaces as a platform for particle self-assembly}
\author{An{\fontencoding{T1}\selectfont\dj}ela \v{S}ari\'c and Angelo Cacciuto}
\affiliation{Department of Chemistry, Columbia University\\ 3000 Broadway, MC 3123\\New York, NY 10027 }
\renewcommand{\today} 

\begin{abstract}
We perform numerical simulations to study self-assembly of nanoparticles mediated by an elastic planar surface. We show how the nontrivial elastic response to deformations of these surfaces leads to anisotropic 
interactions between the particles resulting in aggregates having different geometrical features.
The morphology of the patterns can be controlled by the mechanical properties of the surface and the strength of the particle adhesion. We use simple scaling arguments to understand the formation of the different structures, and we  show how the adhering particles can cause the underlying elastic substrate to wrinkle if two of its opposite edges are clamped.  
Finally, we discuss the implications of our results and suggest how elastic surfaces could be 
used in nanofabrication.
 
\end{abstract}
\maketitle

\section{Introduction}
Elastic surfaces are ubiquitous in nature and technology and appear across all length scales, from the cellular microenvironment to large-scale objects such as bridges and buildings. The mechanical properties of these surfaces play an important role both in their biological function, and in their wide usage in material engineering. For instance, it is known that the stiffness of an elastic substrate alters the morphology and dynamics of tissue cells adhering onto it~\cite{wang}. Variable cytoskeleton assembly~\cite{engler} and cell spreading~\cite{janmey} on substrates of different mechanical properties are two nice examples of this. Furthermore, their response to external stress  have been exploited in metrology~\cite{russell,simony} and in the production of micro- and nano-scale patterned surfaces that may serve as components with novel optical, electronic and magnetic properties~\cite{fery}.
 
We are interested in understanding how 
elastic surfaces could be used to template aggregation of nanocomponents.
The idea of using interfaces, specifically fluid ones, as a means of driving self-assembly of arbitrary building blocks was first introduced by Whitesides et. al.~\cite{whitesides,whitesides2}.
In this case, the presence of floating objects on a fluid interface induces local deformations in its profile which are minimized when the objects are isotropically  
driven  close to each other\cite{isra,gennes}. By manipulating the tension of the interface, and by tailoring the chemistry of the building blocks, millimeter-size objects, microchips and microcomponents have been successfully self-assembled~\cite{whitesides3, hanein,sato}.  
Unlike fluid interfaces,  whose large scale physical properties are dominated by their surface tension, elastic  surfaces resist stress and respond to it in a spring-like fashion.

Elastic surfaces bend and stretch in response to deformations. The resulting macroscopic behavior is characterized by strong nonlinearities~\cite{landau}. The mechanical properties of macroscopic elastic sheets have recently been the subject of intense investigation~\cite{cerda1,cerda2,cerda3,skin,russell,witten}. 
Under an applied force an elastic surface deforms in a  way that minimizes the energy associated with the deformation. It is easy to show~\cite{landau,witten} that the ratio between stretching and bending energies for an arbitrary deformation of amplitude $h$ on a flat elastic sheet of thickness $t$ scales as $E_s/E_b \sim (h/t)^2$. Therefore, for sufficiently thin sheets, bending is the preferred mode of deformation and 
unstretchability  can be thought of as an overall geometrical constraint to the deformations. The net result is that thin elastic surfaces respond to an external applied stress with stretch-free deformation involving (when possible) exclusively uniaxial bending. Skin wrinkling under applied stress~\cite{cerda2,skin} and stress focusing via d-cone formation of crumpled paper~\cite{witten} are two examples of this phenomenon. 

Such nontrivial phenomenology extends to the micro-scale.
There are several artificial and naturally occurring examples of microscopic elastic surfaces, including graphite-oxide sheets~\cite{spector,wen}, graphene sheets~\cite{ruoff,meyer}, cross polymerized biological membranes~\cite{tundo}, cross polymerized 
hydrogels~\cite{janmey}, buckypaper~\cite{endo, hodak,hall}, the spectrin-actin network forming the cytoskeleton of 
Red Blood Cells~\cite{branton,lei}, and very recently they have been fabricated using close-packed nanoparticle arrays~\cite{lin}.
Our expectation is that  diffusible particles adhering over an elastic surface should be 
driven to aggregate into configurations that reduce the mechanical cost of the overall surface deformation. These configurations will depend on the geometry of the surface, its elastic properties and the strength of the adhesion
(the applied force). 

We have recently shown how the response to deformations of elastic nanotubes can alter the elastic properties of a flexible filament binding to it~\cite{saric}, and that elastic nanotubes and nanoshells can drive self assembly of nanoparticles in a variety of patterns that depend on the interplay between bending and stretching rigidities of the template, and the amount of deformation of the surface~\cite{pamies,saric1}.
In this paper we explore the phase behavior of nanoparticles adhering onto a planar (extended) elastic substrate as a function of the mechanical properties of the substrate, namely its stretching and bending rigidity, and the strength of the adhesion. 
We also analyze the role of the boundaries of the elastic sheet and their influence on the aggregation patterns. Our findings suggest that the geometrical features of the anisotropic aggregation of the particles can be tuned in a variety of patterns by controlling the elastic parameters of the problem.

\section{Methods}

The elastic plane is modeled via a standard triangulated mesh with hexagonal symmetry~\cite{mesh}. To impose surface self-avoidance we place hard beads at each node of the mesh.
Any two surface beads  interact via a repulsive truncated-shifted  Lennard-Jones potential:
\begin{equation} \label{LJ}
U_{LJ}=
\begin{cases}
4\epsilon\left[ \left( \dfrac{\sigma}{r}\right)^{12}-\left(\dfrac{\sigma}{r}\right)^{6} + \frac{1}{4}\right] & \text{, $r\leq 2^{1/6} \sigma$}\cr
0 &\text{, $r>2^{1/6} \sigma$}\cr
\end{cases}
\end{equation}
where $r$ is the distance between the centers of two beads, $\sigma$ is their diameter, and
$\epsilon=100 k_{\rm{B}}T$.

We enforce the surface fixed connectivity by linking every bead on the surface to its first neighbors via a harmonic spring potential
\begin{equation} \label{spring}
U_{stretching}=K_s(r-r_B)^2 .
\end{equation}
Here $K_s$ is the spring constant and it models the stretching rigidity of the surface. $r$ is the distance between two neighboring beads, $r_B=1.23\sigma$ is the equilibrium bond length, and it is sufficiently short to prevent overlap between any two triangles on the surface even for moderate values of $K_s$.

The bending rigidity of the elastic surface is modeled by a dihedral potential between  adjacent triangles on the mesh:
\begin{equation} \label{dihedral}
U_{bending}=K_b(1+\cos\phi)
\end{equation}
where $\phi$ is the  dihedral angle between opposite vertices of any two triangles sharing an edge and  $K_b$ is the bending constant.

Particles of diameter $\sigma_c=10\sigma$ are described via 
the repulsive truncated-shifted Lennard-Jones potential
introduced in Eq.~\ref{LJ} with $\sigma\rightarrow\sigma_c$.
The generic binding between the nanoparticles and surface is described by a Morse potential:
\begin{equation} \label{morse}
U_{Morse}=
\begin{cases}
D_0\left( e^{-2\alpha(r-r_{NB})}-2e^{-\alpha(r-r_{NB})}\right) & \text{, $r\leq 10 \sigma$}\cr
0 &\text{, $r>10 \sigma$}\cr
\end{cases}
\end{equation}
where $r$ is the center-to-center distance between a nanoparticle and a surface-bead,  $r_{NB}$ is bead-nanoparticle contact distance $r_{NB}=5.5\sigma$ and
$D_0$ is the binding energy. The interaction cutoff is set to 10$\sigma$ and $\gamma=1.25/\sigma$.   
 
The simulations were carried out using the {\sc LAMMPS} molecular dynamics package~\cite{lammps} with a Langevin dynamics in the $NVT$ ensemble. Dimensionless units are used throughout this paper. The timestep size was set to $dt=0.002\tau_0$ ($\tau_0$ is the dimensionless time) and each simulation was run for a minimum of $5 \cdot 10^{6}$ iterations. In this study we considered unconstrained and edge-constrained sheets.
To minimize edge effects in unconstrained sheets we considered surfaces with 
an overall circular geometry. Two different equilibrium radii 
$R_{plane}\simeq50\sigma$ and $R_{plane}\simeq60.4\sigma$ were explored. To preserve the mechanical stability of the sheet the colloids were placed both on top and at bottom of the surface.
When edge-constrained surfaces were considered, a rectangular shape was selected and the colloids were placed only on one side of the plane.
For this specific case we considered two surface equilibrium areas, $A\simeq(176 \times 152)\sigma^2$ and $A\simeq(244\times 212)\sigma^2$. In both cases a wide range of nanoparticle surface fractions between $10\%$ and $60\%$ was explored.
Typical values of $\sigma\sim 10-20 nm$ would  imply colloidal particles of diameter $\sim$ 100-200 nm and surfaces of
area $A\sim 200-1000 \mu {\rm m} ^2$. Figure~\ref{fig1} illustrates the model used in our simulations. 

\section{Results}
\begin{figure*}
\includegraphics[width=60mm]{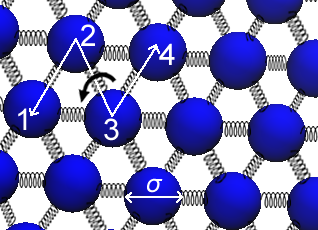}
\caption{Illustration of the triangulated mesh model used in our simulations. The surface beads of diameter $\sigma$ (blue spheres) are set at the nodes of each triangular element to enforce surface-self-avoidance and are linked to their first neighbors with springs of the constant $K_s$ and the equilibrium length $r_{NB}$ (measured form the beads centres). 
The surface connectivity is kept constant, and apart from boundary nodes each surface bead has six neighbors. 
The dihedral angle 1-2-3-4 from which bending energies are computed is also indicated. This energy is minimized when all angles between neighboring triangles are equal to $\phi$.}
\label{fig1} 
\end{figure*}

We find that elastic surfaces can drive nanoparticle aggregation. The geometry of the 
aggregates can be tuned into a variety of patterns   controlled by the 
mechanical properties of the surface ($K_s$ and $K_b$) and the strength of the particle's adhesion ($D_0$). Let us begin by looking at the role of the membrane's stretching rigidity. Fig.~\ref{fig1}a) shows a diagram of the different aggregates obtained for
different values of $K_s$ as a function of the extent of the surface deformation (regulated by $D_0$) at fixed bending rigidity $K_b$. Fig.~\ref{fig1}c) shows simulation snapshots of the corresponding patterns. As expected, when $D_0$ is small the surface is basically unaffected by the presence of the particles and the particles behave effectively as a low-density two-dimensional hard sphere fluid. 
In the opposite limit, when the particles bind very strongly, the membrane undergoes large local deformations limiting the diffusion of the particles and resulting in kinetically trapped configurations. Repeating the simulations under the same conditions leads to a different not well defined configuration. We call this phase the arrested phase. 
 
The intermediate regime is characterized by five distinct structured phases. 
For small values of $K_s$ the aggregation is completely driven by the minimization of the bending energy. As a result particles  aggregate into a two-dimensional hexagonal crystal. This is what happens for instance in lipid bilayers where a bending-mediated
isotropic, $V\sim-r^{-4}$, interaction can drive surface inclusions at a separation $r$ from each other to aggregate~\cite{goulian}.
Upon a small increase in $K_s$ the hexagonal crystal rearranges into a square lattice. 
We believe the coexistence of two lattices is either due the substrate having the elastic modulus near the boundary between the hexagonal and square crystalline phases, or due to small spatial inhomogeneity of the substrate's elastic modulus, either way, this recent experimental observation seem compatible with our numerical results. As $K_s$ is further increased we find that the crystalline aggregate is disrupted in favor of a network of short connected lines. 
This phase is the consequence of balancing the stretching and the bending energy: the former preferring the stretch-free uniaxial deformations, and the latter driving in-plane isotropic aggregation.

For even larger values of  $K_s$ the connected network is disrupted and particles 
arrange into straight parallel lines.  Increasing $K_s$ at this points only leads to
 to a larger stiffness of the linear aggregates.
This transition is completely stretching-driven. The parallel lines start appearing when $K_s\gtrsim K_b$.  This is clearly shown in Fig.~\ref{fig1}b) where we show how the formation of straight and connected aggregates depend on both stretching and bending constants. For $K_s/K_b\gg1$ one indeed recovers the  thin and unstreachable sheet limit for which only stretch-free (uniaxial) deformations are possible. The most dramatic consequence of this property of elastic plates is the fifth, folded phase. This phase occurs for larger values of $D_0$,  when particles tend to increase the contact area with the membrane as much as possible. In this region the surface immediately folds 
into a well organized higher three-dimensional hexagonal structure (Fig.~\ref{fig1}c)).
 
\begin{figure*}
\includegraphics[width=160mm]{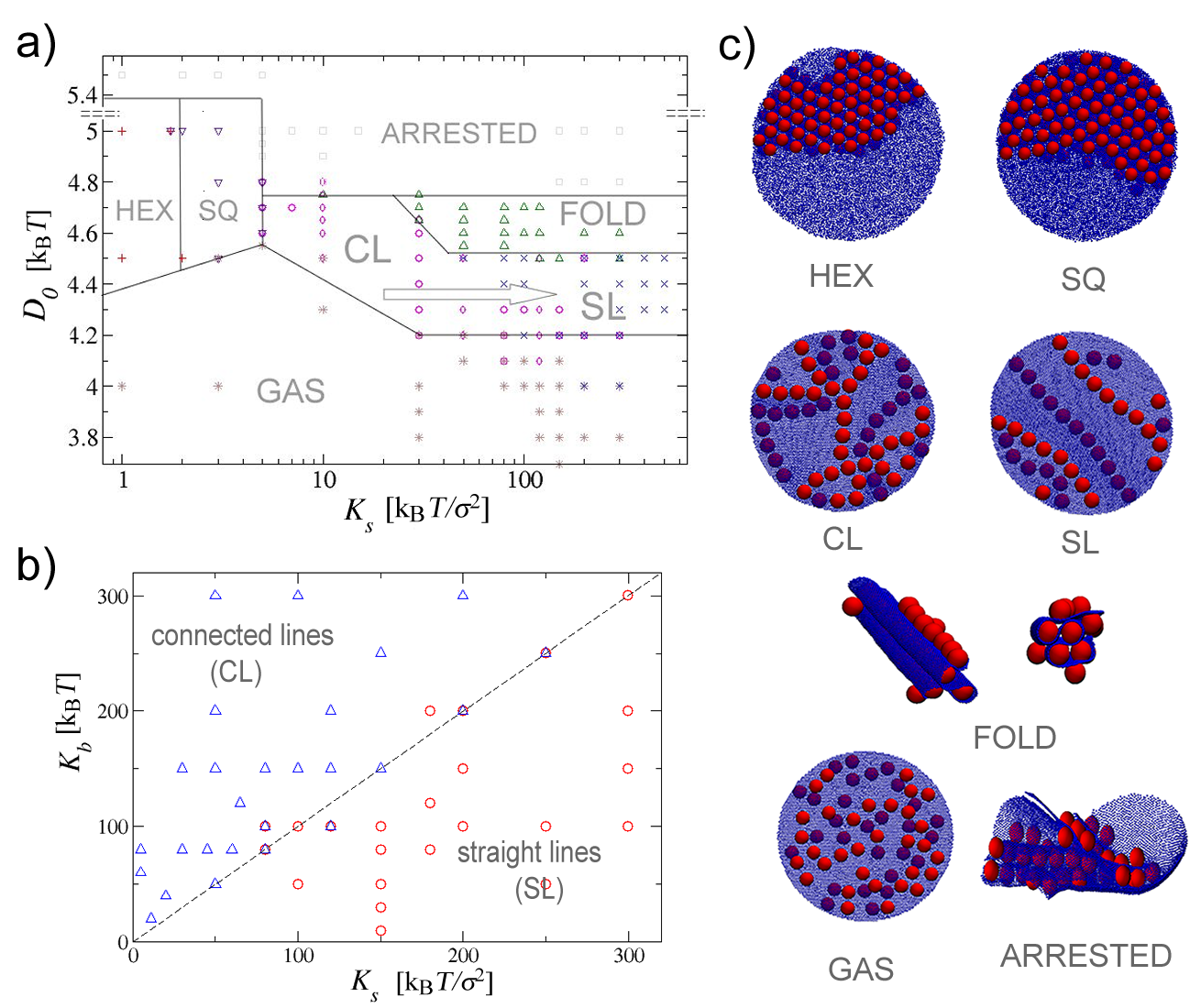}
\caption{(a) Phase diagram of nanoparticles binding to an elastic planar surface. In this case 
the bending rigidity is $K_b=150k_{\rm B}T$; the equilibrium radius of the surface 
is $R\simeq 60.4 \sigma$ and the number of nanoparticles is $N=40$. 
The lines separating the different phases serve as a guide to the eye. The arrow points in the direction of lower line  connectivity. (b) Boundary between connected to straight parallel lines as a function of $K_s$ and $K_b$. The dashed $K_b=K_s$ line serves as a guide to the eye. (c) Simulation snapshots of the seven observed phases. For the sake of clarity the hexagonal and the square crystal phases are shown with the larger number of particles than the other phases.}
\label{fig2} 
\end{figure*}

To better characterize the dependence of the different phases on $K_s$ - from the hexagonal to the square lattice, from the connected network to the linear one, 
we also measured the frequency of particle contacts as a function of $K_s$. 
Fig.~\ref{fig2} shows the probability distribution  
of the number of the nearest neighbors as a function of $K_s$ in the different phases. 
The connectivity decreases when increasing  $K_s$, going 
from the six neighbors of the hexagonal phase to the four neighbors in the square lattice, and finally two for the connected and the straight lines. For the linear aggregates the significant difference is not in the location of the peak of the distribution (indeed a large number of particles will have two neighbors even in the connected linear aggregates), but in the relative height of $P(3)$ which is the signature for branching.
 
It should be emphasized that the number of connections does decrease continuously  with
increasing $K_s$. It is tempting to interpret these data in terms of a single growing length scale that sets the size for the average distance between any two nodes in the linear network, and 
consider the straight-line phase as the limiting behavior in which this distance becomes larger than the surface. A simple mean field calculation balancing stretching and bending energies~\cite{witten} points to the length scale $l_p\propto h^{1/2}(K_s/K_b)^{1/4} R^{1/2}$, which qualitatively produces  the correct 
phenomenological behavior, but unfortunately the small system sizes analyzed in this study prevent us from making such a link  more concrete.

\begin{figure*}
\includegraphics[width=100mm]{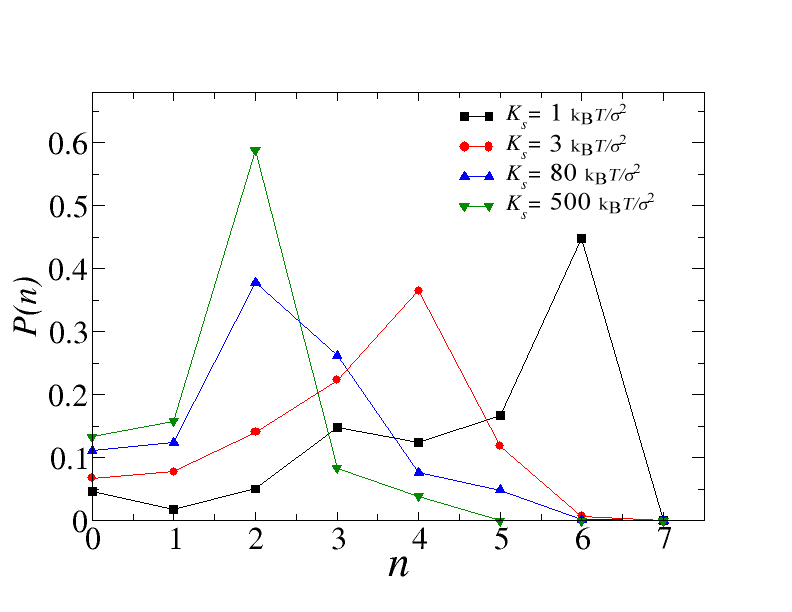}
\caption{Probability distributions of particle contacts in a self-assembled aggregates  for different values of  $K_s$, and constant $K_b=150k_{\rm B}T$. From right to left the distributions refer to the planar hexagonal crystal, the planar square crystal, the interconnected   lines,  and the straight parallel lines.}
\label{fig3}   
\end{figure*}

It is important to stress that the free boundaries of the membrane play an important role.
Indeed, it is not clear whether the linear phases indicated in Fig.~\ref{fig1} are stable with respect to folding. In fact, in  a few cases our longest simulations of the linear phases resulted eventually in a folded phase. We expect this to be an effect of the free boundary of the surface that can be taken care of by applying a small external tension or by clamping the outer edge of the surface.
To show that this is indeed the case, we considered a rectangular elastic sheet 
in which two opposite sides (edges) are kept fixed (clamped). 
In the absence of the adhering particles the sheet remains flat to its equilibrium 
size. Once the particles bind to it 
we observe only two phases for moderate values of $K_s$,
the gas phase and the straight-linear phase. The former appears when $D_0$ is insufficient for the particles to deform the membrane, while the latter  occurs when $D_0$ crosses a certain threshold value which depends mainly on the bending rigidity of the plane (Fig.~\ref{fig3}a)).\

The linear structures formed in this phase always appear to be perpendicular to the constrained sides of the membrane (Fig.~\ref{fig3}b)). However we observe that the distance between them can be tuned by changing $K_s$ and $K_b$. This kind of pattern is reminiscent of the wrinkle pattern that occurs when a thin elastic sheet is subjected to a longitudinal stretching strain\cite{cerda2,cerda3}. The sheet is then unable to contract laterally near the clamped boundaries, so it wrinkles to accommodate the in plane stress. Cerda and Mahadevan showed that, for a constant tension, the wavelength of the wrinkles scales as $\lambda\sim(L)^{1/2}\sim(K_b/K_s)^{1/4}$ \cite{cerda3}.\\
Indeed, we find the same reasoning can be applied here. Instead of having an external force stretching the plane, the particles  binding to the surface act as the stress source causing the sheet to wrinkle. This stress is directed perpendicularly to the fixed sides of the plane. Since the wrinkles are the regions where the particles can gain the highest contact area with the surface, i.e the highest binding, the particles follow the wrinkle pattern resulting in the straight parallel aggregates. 
We believe that the destabilization of the linear-connected phase 
is due to the implicit symmetry breaking imposed by the way we clamp the membrane,
in fact, when clamping is enforced on all four edges of the sheet, the phase reappears. 

We also analyzed the dependence of the wavelength of the particles' lines with $K_b$ and $K_s$ and it appears to nicely follow the theoretical prediction of Cerda and Mahadevan (Fig.~\ref{fig3}b)). Nevertheless, two extra parameters play a role in determining the separation between the lines in this case: the surface coverage and the particles's binding energy. Since particle binding to the surface is favorable, once the particle's density becomes larger than that required to completely 
fill the wrinkles with particles,  new lines (wrinkles) form in between the preexisting ones, bringing the preexisting ones closer together. The inset in Fig.~\ref{fig3}b) shows the decrease in $\lambda$ with the increase in the particle density, for two different values of $K_s$. In addition to that, we find that the increase in $D_0$ (for constant $K_s$ and $K_b$) also brings the lines closer together. Higher binding increases the amplitude of the wrinkles (analogous to increasing the strain tension in \cite{cerda2}), which decreases the surface area accessible to the particles, effectively increasing the density. 

It should be stressed that the mechanism driving self-assembly of particles into linear aggregates that 
we described is significantly different from  the controlled wrinkling methods recently
developed for the fabrication of patterned surfaces\cite{fery}. 
There the wrinkles are preformed by compressing the substrate, and particles trivially 
arrange along the wrinkles' axis to maximize their binding energy, in our case the surface is not 
pre-wrinkled, and the linear aggregates develop (in a reversible manner)
as a result of a more delicate balance between the  energies of the system and the collective behavior of the particles. Interestingly, once the wrinkled phase is formed it is possible to control the
overall direction of the lines by simply applying a small external tension. 
For instance, the release of  the surface clamping and simultaneous application of a small tension 
in the direction perpendicular to the direction of the wrinkles, results in a reorientation of the lines
along the direction of the tension. This supports our assumption of the reversibility of the line-forming process and could suggests even richer potential application of this approach for periodical patterning.\\

\section*{CONCLUSIONS}
In this paper we shown how elastic surfaces can template self-assembly of nanoparticles,
similarly to the way fluid interfaces do. We show how by tuning the relative cost of bending and stretching
energies (i.e. the thickness of the sheet) it is possible to control the geometry of the aggregates.
The formation of the different linear aggregates, for thin sheets, is an explicit  manifestation of the 
anisotropic interaction between the nanoparticles. 
When the surfaces become effectively unstreachable
particles arrange into macroscopic ordered parallel lines whose separation can be controlled 
by the elastic parameters of the surface. Clamping of the edges across the membrane substantially improves the periodic ordering in the system.

The physical properties of our model can be mapped onto a model of a thin 
sheet supported on an elastic foundation if the stretching rigidity of the plane is 
substituted by the stiffness of the elastic foundation.
Therefore, the results of our  theoretical study are quite general and may suggest
novel use of the elastic interfaces in nano/micromechanics and material engineering. Possible experimental systems where our predictions could be tested include cross-polymerized or crystalline lipid bilayers,
thin polymeric sheets, ultrathin cross-linked nanoparticle-membranes  or possibly free standing liquid crystalline films in the presence of colloidal particles. More in general, any elastic substrate that can be locally deformed by the interaction with 
a diffusable binding component.

\section*{ACKNOWLEDGMENTS}
This work was supported by the National Science Foundation under Career Grant No. DMR-0846426. We thank Josep C. P\`amies and William L. Miller for helpful discussions.

\begin{figure*}
\includegraphics[width=160mm]{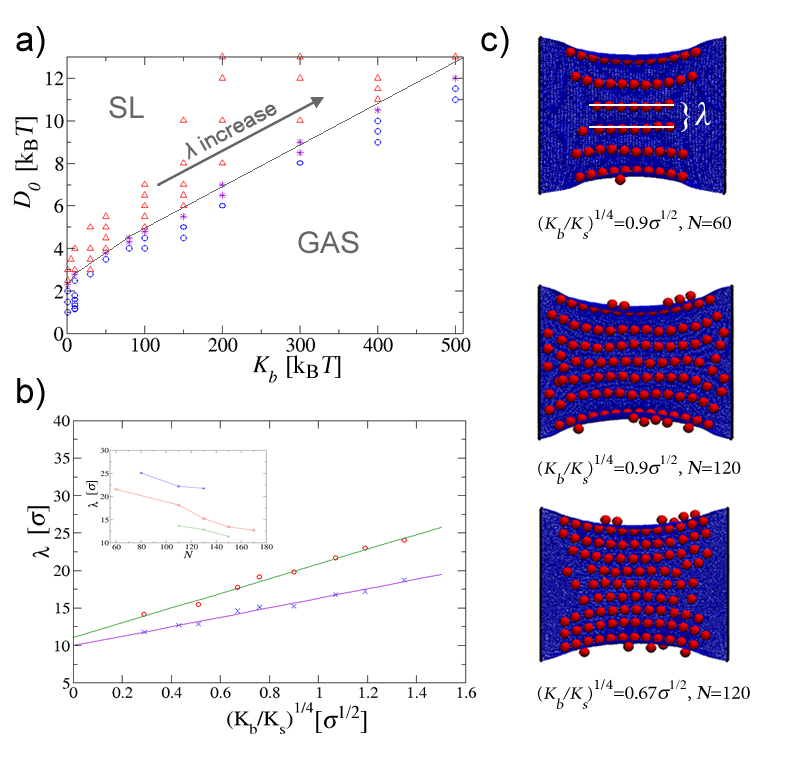}
\caption{(a) Phase diagram of nanoparticles binding to a clamped rectangular elastic surface. Two phases are observed  for different
values of  $K_b$ and depend only on $D_0$: the gas phase and straight parallel lines phase (SL). 
These data refer to the case in which $K_s=150 k_{\rm B}T/\sigma^2$, $N=60$,  and the area of the the plane is 
$(176\times152) \sigma^2$. (b) Line separation $\lambda$ as a function of the mechanical properties of the surface.
We show data for two different surface coverages: $\phi=38.7\%$ (cross symbols) and $\phi=21.2\%$ (circle symbols). The straight lines represent the fit of the data to the scaling law $\lambda\sim(K_b/K_s)^{1/4}$. The inset shows the dependence of $\lambda$ on the particle surface coverage, shown for three different values of the $(K_b/K_s)^{1/4}$ parameter: 1.35$\sigma^{1/2}$(top), 0.9$\sigma^{1/2}$ (middle), 0.51$\sigma^{1/2}$ (bottom). (c) Simulation snapshots of linear aggregates for three different combinations of the elastic parameters and the surface coverage densities. The equilibrium surface area is $A=(176\times152) \sigma^2$.  }
\label{fig4}  
\end{figure*}
\end{document}